# Lookup Table-Based Consensus Algorithm for Real-Time Longitudinal Motion Control of Connected and Automated Vehicles

Ziran Wang[1], *Student Member*, *IEEE*, Kyungtae Han[2], *Senior Member*, *IEEE*, BaekGyu Kim[2], *Member*, *IEEE*, Guoyuan Wu[1], *Senior Member*, *IEEE*, and Matthew J. Barth[1], *Fellow*, *IEEE*

*Abstract*—Connected and automated vehicle (CAV) technology is one of the promising solutions to addressing the safety, mobility and sustainability issues of our current transportation systems. Specifically, the control algorithm plays an important role in a CAV system, since it executes the commands generated by former steps, such as communication, perception, and planning. In this study, we propose a consensus algorithm to control the longitudinal motion of CAVs in real time. Different from previous studies in this field where control gains of the consensus algorithm are pre-determined and fixed, we develop algorithms to build up a lookup table, searching for the ideal control gains with respect to different initial conditions of CAVs in real time. Numerical simulation shows that, the proposed lookup table-based consensus algorithm outperforms the authors' previous work, as well as van Arem's linear feedback-based longitudinal motion control algorithm in all four different scenarios with various initial conditions of CAVs, in terms of convergence time and maximum jerk of the simulation run.

## I. Introduction

Significant developments of connected and automated vehicle (CAV) technology have been achieved during the last decade, where vehicles not only are enabled to be driven autonomously by themselves, but also can cooperate with each other through vehicle-to-vehicle (V2V) communications. Besides communication, perception, and planning, control also plays a crucial role in the architecture of a CAV, since it is the execution part of the architecture and it greatly affects the safety and efficiency of a CAV. The control system of a CAV can be categorized into two levels: The higher level is the motion control, which regulates the longitudinal and lateral motion of the vehicle and ensures the reference behavior is executed in the closed loop; The lower level is the powertrain control, which controls the power generation for vehicle motion and auxiliary loads, and it governs the operation of many components such as transmissions, internal combustion engines, electric motors, starters, and generators and so on [1].

Longitudinal motion control of CAVs has been extensively studied by researchers around the world, where the distributed consensus algorithm is considered one of the real-time approach. Essentially, consensus means a network of agents cooperatively reach an agreement with respect to a certain interest which depends on the states of all agents. Instead of being controlled by a centralized scheme, which assumes the global team knowledge is available to all agents in the network, consensus can act as a distributed scheme which requires only local interactions and evolves in a parallel manner [2]-[5].

Cooperative adaptive cruise control (CACC) is a typical application of CAV technology where consensus algorithm can be designed and implemented. A CACC system consists of a string of vehicles which take advantage of V2V communications to follow each other with harmonized speed and headway [6]. Specifically, single-integrator distributed consensus algorithms were developed to achieve weighted and constrained consensus for inter-vehicle distance in the vehicle string [7], [8]. Double-integrator distributed consensus algorithms were further developed based on single-integrator, where different information flow topologies of vehicles in the system were adopted by different work, such as predecessor following and predecessor-leader following [9]-[12]. The CACC technology using consensus algorithm was also applied to other use cases like cooperative on-ramp merging, where a vehicle can be projected on the other lane as a "ghost" leader for the follower to follow through V2V communications [13], [14]. Compared to model predictive control (MPC), which is another dominant motion control approach for CACC, the consensus algorithm or other linear feedback algorithms have the advantages of less computational load and the easiness to implement on real vehicles [15].

Although the stability and robustness of the consensus algorithm in the use of longitudinal motion control were elaborately discussed and analyzed in aforementioned work, none of them focused on validating the system performance in terms of real-world constraints. For example, how to tune the control gains of the consensus algorithm to satisfy the safety constraint, efficiency constraint and comfort constraint was not covered. Different initial states of vehicles (e.g., longitudinal position, speed, etc.) have a huge impact on the performance of the consensus algorithm, so how to adjust the values of control gains to allow the consensus algorithm to work well for different scenarios remains a critical issue. However, solving an optimal problem with multiple constraints takes a relatively long time, and it is not feasible to do that when the consensus algorithm is running at every time step and in every scenario. Therefore, if we run the simulations with all possible initial states offline, and build up a lookup table to store the ideal control gains, then when the algorithm is running in real time, it can simply search for the ideal control gains in the table based on its conditions. In such a manner, the computational load of the proposed algorithm will be dramatically decreased, making the algorithm a real-time control algorithm for vehicles.

[1] Ziran Wang, Guoyuan Wu and Matthew J. Barth are with the College of Engineering – Center for Environmental Research and Technology (CE-CERT), University of California, Riverside, CA 92507, USA. Email: zwang050@ucr.edu, gywu@cert.ucr.edu, barth@ece.ucr.edu.

[2] Kyungtae Han and BaekGyu Kim are with Toyota InfoTechnology Center, Mountain View, CA 94043, USA. Email: kthan@us.toyota-itc.com, bkim@us.toyota-itc.com.

The main contribution of the paper is to develop a consensus algorithm for real-time motion control of CAVs based on the authors' prior study [13]. Specifically, we develop algorithms to build up a lookup table of the consensus algorithm's control gains, searching the control gain which satisfies all system constraints with respect to different initial states of vehicles. When the vehicle adopts consensus algorithm in applications like CACC or cooperative on-ramp merging, it selects the ideal values of control gains based on the states of itself and its leader. With searching the control gains in the lookup table generated offline, the consensus algorithm can be run in real time to control the longitudinal motion of a CAV.

The remainder of this paper is organized as follows: Section II describes the problem to solve in this study, and highlights its differences with respect to existing studies. Section III proposes the consensus algorithm, and analyze its convergence and string stability. Section IV develops algorithms to build up lookup table for the control gains, satisfying the constraints of safety, efficiency and comfort. Simulation study is conducted given different initial states, and results are shown in section V. Section VI concludes the paper with some further directions.

## II. PROBLEM STATEMENT

Given the second-order dynamics of a vehicle $i$

$$\dot{r}_i(t) = v_i(t) \quad (1)$$

$$\dot{v}_i(t) = a_i(t) \quad (2)$$

where $r_i(t)$, $v_i(t)$ and $a_i(t)$ denote the longitudinal position, longitudinal speed and longitudinal acceleration of vehicle $i$ at time $t$, respectively. In this study, we propose the consensus algorithm based on the predecessor following information flow topology of a string of vehicles, where the following vehicle only gets information from its immediate leading vehicle through V2V communications. Given that, the problem can be generalized as a car-following problem shown as Fig. 1.

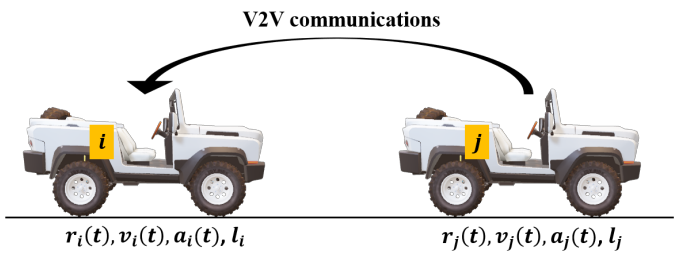

Fig. 1. Illustration of the predecessor following information flow topology.

In this figure, the term $l_j$ denotes the length of vehicle $j$. As can be seen, the following vehicle $i$ receives information from the leading vehicle $j$ through V2V communications. Therefore, the problem of forming a predecessor following string of vehicles can be formulated as that, given determined $l_i$ and $l_j$, and $r_i(0), v_i(0), a_i(0), r_j(0), v_j(0), a_j(0)$, how to apply a longitudinal control algorithm such that

$$r_i(t) \rightarrow r_j(t) - r_{headway} \quad (3)$$

$$v_i(t) \rightarrow v_j(t) \quad (4)$$

$$a_i(t) \rightarrow a_j(t) \quad (5)$$

where "$\rightarrow$" means the value on the left-hand side converges to the value on the right-hand side, $r_{headway}$ is the desired distance headway between two vehicles.

Most existing relevant works adopt the same set of control gains (also known as damping gains) independently of certain driving scenario characteristics. Such uniform assignment of control gains may not guarantee the constraints of the proposed consensus algorithms under some driving scenarios. For example, the consensus algorithm might work well when the initial speed difference of two vehicles are relatively small, and the initial headway is relatively large. However, when the initial conditions of these parameters change dramatically but the control gains remain the same, some overshoot of the headway might appear during the convergence process. Although the dynamics of two vehicles will still converge to consensus eventually, safety constrains cannot be satisfied since rear-end collision between vehicles will happen. The convergence to consensus might also take a very long time, making the algorithm inefficient. Additionally, the speed or acceleration can change dramatically during a short period without any comfort constraint on the consensus algorithm. These are the issues we want to address in this study by building up a lookup table of control gains.

## III. CONSENSUS ALGORITHM

### A. Preliminaries and Background

The information flow topology of a string of vehicles can be represented by a directed graph $\mathcal{G} = (\mathcal{H}, \mathcal{E})$, where $\mathcal{H} = \{1, 2, ..., n\}$ is a finite nonempty node set and $\mathcal{E} \subseteq \mathcal{H} \times \mathcal{H}$ is an edge set of ordered pairs of nodes (i.e., edges). The edge $(i, j) \in \mathcal{E}$ denotes that vehicle $j$ can obtain information from vehicle $i$. It is not necessarily true vice versa, since the information flow is not always bi-directional. The neighbors of vehicle $i$ are denoted by $\mathcal{N}_i = \{j \in \mathcal{H}: (i, j) \in \mathcal{E}\}$. The topology of the graph is associated with an adjacency matrix $\mathcal{A} = [a_{ij}] \in \mathbb{R}$, which is defined such that $a_{ij} = 1$ if edge $(j, i) \in \mathcal{E}$, $a_{ij} = 0$ if edge $(j, i) \notin \mathcal{E}$, and $a_{ii} = 0$. $\mathcal{L} = [\ell_{ij}] \in \mathbb{R}$ (i.e., $\ell_{ij} = -a_{ij}$, $i \neq j$, $\ell_{ii} = \sum_{j=1, j \neq i}^{n} a_{ij}$) is the nonsymmetrical Laplacian matrix associated with $\mathcal{G}$. A directed spanning tree is a directed tree formed by graph edges that connects all the nodes of the graph.

Before proceeding to develop our consensus algorithm for longitudinal motion controller, we recall here the fundamental consensus algorithms which lay the foundation of our study. Given the continuous communication between two agents in the distributed networks, a differential equation can be proposed to model the information state update of each agent. The double-integrator distributed consensus algorithm [16] can be given by

$$\dot{r}_i(t) = v_i(t) \quad (6)$$

$$\dot{v}_i(t) = -\sum_{j=1}^{n} a_{ij} k_{ij} \cdot [(r_i(t) - r_j(t)) + \gamma \cdot (v_i(t) - v_j(t))], \, i \in \mathcal{H} \quad (7)$$

where $k_{ij} > 0$ is a control gain, and $\gamma > 0$ is another control gain that denotes the coupling strength between the state derivatives. Consensus of this algorithm is reached when $r_i(t) \to r_j(t)$ and $v_i(t) \to v_j(t)$.

### B. Longitudinal Motion Controller for CAVs

The double-integrator distributed consensus system is a good approach to tackle formation control problems of intelligent agents such as CAVs. There are a few points that we need to be aware of when applying the above algorithm to a car-following problem like Fig. 1. In equation (7), the desired position between two agents is zero, i.e., $r_i(t) - r_j(t) \to 0$. However, in a car-following problem like Fig. 1, we put $(r_i(t) - r_j(t))$ in the position consensus part to set the desired distance headway between two vehicles. In addition, time delay is not considered in equation (7), while communication delay is not neglectable in the car-following case of CAVs. Therefore, the longitudinal motion controller of CAVs in a car-following case is proposed as

$$\dot{r}_i(t) = v_i(t) \quad (8)$$

$$\dot{v}_i(t) = -a_{ij}k_{ij} \cdot \left[\left(r_i(t) - r_j\left(t - \tau_{ij}(t)\right) + l_j + v_i(t) \cdot \left(t_{ij}^g(t) + \tau_{ij}(t)\right)\right) + \gamma_i \cdot \left(v_i(t) - v_j\left(t - \tau_{ij}(t)\right)\right)\right], i,j \in \mathcal{H} \quad (9)$$

where $\tau_{ij}(t)$ denotes the time-variant communication delay between two vehicles, $t_{ij}^g(t)$ is the time-variant desired time gap between two vehicles, which can be adjusted by many factors like road grade, vehicle mass, braking ability, etc.

If we define $\tilde{r}_i$ and $\tilde{v}_i$ as the position error and speed error of vehicle $i$ with respect to the leading vehicle of a vehicle string (i.e., the desired values of all following vehicles in the string), then equation (8) and (9) can be rewritten as

$$\dot{r}_i(t) = v_i(t) \quad (10)$$

$$\dot{v}_i(t) = -a_{ij}k_{ij} \cdot [(\tilde{r}_i(t) - \tilde{r}_j(t - \tau_{ij}(t))) + \gamma_i \cdot (\tilde{v}_i(t) - \tilde{v}_j(t - \tau_{ij}(t)))], \, i,j \in \mathcal{H} \quad (11)$$

If we further define the dynamics of the vehicle in a compact form as

$$\tilde{r} = [\tilde{r}_1^T, \tilde{r}_2^T, \ldots, \tilde{r}_i^T, \ldots, \tilde{r}_n^T]^T \quad (12)$$

$$\tilde{v} = [\tilde{v}_1^T, \tilde{v}_2^T, \ldots, \tilde{v}_i^T, \ldots, \tilde{v}_n^T]^T \quad (13)$$

then the state vector can be defined as

$$\tilde{\chi} = [\tilde{r}^T \, \tilde{v}^T]^T \quad (14)$$

The double-integrator vehicle dynamics in equation (10) and (11) can be further transformed into a compact form as

$$\dot{\tilde{\chi}}(t) = \Gamma_1 \tilde{\chi}(t) + \Gamma_k \tilde{\chi}(t - \tau_k(t)) \quad (15)$$

$$\Gamma_1 = \begin{bmatrix} 0_{n \times n} & I_{n \times n} \\ -\tilde{A} & -\gamma \tilde{A} \end{bmatrix}, \Gamma_k = \begin{bmatrix} 0_{n \times n} & 0_{n \times n} \\ \tilde{A}_k & \gamma \tilde{A}_k \end{bmatrix} \quad (16)$$

$$\tilde{A} = \text{diag}\{a_{12}, a_{23}, \ldots, a_{ij}, \ldots, a_{(n-1)n}\} \quad (17)$$

where $\tau_k(t)$, $k = 1, 2, \ldots, m$ with $m \leq n(n-1)$ is defined as an element of the time-varying communication delay $\tau_{ij}(t)$.

Given the Leibniz-Newton formula

$$\tilde{\chi}(t - \tau_k(t)) = \tilde{\chi}(t) - \int_{-\tau_k(t)}^{0} \dot{\tilde{\chi}}(t + s) ds$$

$$= \tilde{\chi}(t) - \Gamma_l \int_{-\tau_k(t)}^{0} \tilde{\chi}(t + s - \tau_l(t + s)) ds \quad (18)$$

where $l = 0, 1, 2, \ldots, m$, and substitute this into equation (15)

$$\dot{\tilde{\chi}}(t) = B\tilde{\chi}(t) - \Gamma_k \Gamma_l \int_{-\tau_k(t)}^{0} \tilde{\chi}(t + s - \tau_l(t + s)) ds \quad (19)$$

$$B = \Gamma_1 + \Gamma_k = \begin{bmatrix} 0_{N \times N} & I_{N \times N} \\ -\bar{A} & -\gamma \bar{A} \end{bmatrix} \quad (20)$$

where $\bar{A} = -\tilde{A} + \tilde{A}_k$.

*Theorem 1*: If there exists a directed spanning tree in the platoon information flow topology $\mathcal{G}$, and the control gain $\gamma$ of equation (9) suffices

$$\gamma_i > \max_{\mu_i \in \eta(\bar{A})} \left\{ \frac{|\text{Im}\{\mu_i\}|}{\sqrt{|\text{Re}\{\mu_i\}| \cdot |\mu_i|}} \right\} \quad (21)$$

where $\mu_i$ is the $i$th eigenvalue of $\bar{A}$, and $\eta(\bar{A})$ is the set of all eigenvalues of $\bar{A}$, then there exists a constant $\tau_0 > 0$ such that when $0 \leq \tau_k \leq \tau_0$ ($k = 1, 2, \ldots, m$), the vehicles in the same string can achieve consensus. Due to the limitation of space in this paper, the proof of this theorem is not covered in this section, but it can be referred to the authors' prior work [9].

String stability is a basic requirement to ensure the safety of a vehicle string like CACC system. Specifically, string stability is a desirable characteristic for vehicle strings to attenuate either distance error, velocity or acceleration along upstream direction. If we consider vehicle $i$ as a following vehicle $j$ ($= i - 1$), then we can write equation (9) in the Laplace domain with time-variant communication delay set to a constant value $\tau$ as

$$A_i(s) = -a_{i(i-1)} k_{i(i-1)} \cdot \left[\left(R_i(s) - R_{(i-1)}(s) e^{-\tau s} + \frac{l_{i-1}}{s} + V_i(s) e^{-\tau s} \frac{(t_{i(i-1)}^g + \tau)}{s}\right) + \gamma_i (V_i(s) - V_{i-1}(s) e^{-\tau s})\right], i \in \mathcal{H} \quad (22)$$

After some algebraic manipulations when assuming low frequency condition, the following equation can be derived

$$\frac{A_i(s)}{A_{i-1}(s)} = \frac{a_{i(i-1)} k_{i(i-1)} \cdot [e^{-\tau s} + se^{-\tau s}(t_{ij}^g + \tau) b_i + s\gamma_i e^{-\tau s}]}{s^2 + \gamma_i s + 1} \quad (23)$$

where the control gains $k$ and $\gamma$ in equation (9) can be chosen to guarantee $\frac{A_i(s)}{A_{i-1}(s)} \leq 1$ and hence satisfy the string stability for all frequencies of interest [17].

## IV. LOOKUP TABLE FOR CONTROL GAINS

Since most existing consensus algorithms only adopt one initial condition of vehicles in the simulation study, one single set of well-defined control gains $k$ and $\gamma$ worked well under that condition. However, given different initial states of CAVs, the consensus algorithm tends to behave differently in terms of overshoot, convergence rate and maximum changing rate. A set of control gains working well under one initial condition does not necessarily mean working well under all other initial conditions. How to find the ideal value of control gains in real time when the initial conditions of vehicles are dynamically changing remains an unsolved problem.

In this part, we propose the algorithms to build up a lookup table, aiming to find the ideal values of control gains in terms of different initial states of the leading vehicle and the following vehicle. Given the second-order dynamics of vehicles as in equations (1) and (2), the initial condition of the following vehicle $i$ and the leading vehicle $j$ are $(r_i(t_0), v_i(t_0), a_i(t_0))$ and $\left(r_j\left(t_0 - \tau_{ij}(t_0)\right), v_j\left(t_0 - \tau_{ij}(t_0)\right), a_j\left(t_0 - \tau_{ij}(t_0)\right)\right)$, where $t_0$ denotes the initial time step when the consensus algorithm is applied. Since the proposed double-integrator consensus algorithm (9) does not consider the acceleration of the leading vehicle, and the positions of vehicles are only calculated as their difference, the initial condition of the proposed consensus algorithm can be simplified to $\left(\Delta r_{ij}(t_0), v_i(t_0), v_j\left(t_0 - \tau_{ij}(t_0)\right)\right)$, where $\Delta r_{ij}(t_0) = r_j\left(t_0 - \tau_{ij}(t_0)\right) - r_i(t_0)$. The reason we cannot simplify the initial speed of two vehicles into one term is that the term $v_i(t_0)$ is also calculated in the position consensus term (10), so the value of each vehicle's speed matters to the consensus algorithm.

Every time the consensus algorithm (9) starts to run on the vehicle $i$, the value of control gains $k_{ij}$ and $\gamma$ can be set in real time with the initial condition of vehicles $\left(\Delta r_{ij}(t_0), v_i(t_0), v_j\left(t_0 - \tau_{ij}(t_0)\right)\right)$. The method is to build a 3-dimension lookup table offline covering certain possible values of the initial conditions within certain sets, and the ideal values of control gains can be picked from certain sets of candidates. The three major constraints we consider when choose the ideal control gains are: safety, efficiency, and comfort.

*Constraint 1: Safety Constraint.* The overshoot of the algorithm influences the safety of the longitudinal motion controller. Since the consensus algorithm is proposed to control the longitudinal motion of vehicles, overshoot of the longitudinal position might cause rear-end collision between two vehicles. Therefore, the following constraint should be satisfied to guarantee the safety of the longitudinal motion controller

$$r_j\left(t - \tau_{ij}(t)\right) - r_i(t) > l_j, t \in [t_0, t_{consensus}] \quad (24)$$

where $t_{consensus}$ denotes the time step when consensus is reached. If the headway between the leading vehicle and the following vehicle is no greater than the length of the leading vehicle, a rear-end collision happens. Control gains should be set to guarantee no overshoot of the headway.

*Constraint 2: Efficiency Constraint.* The convergence rate of the consensus algorithm influences the efficiency of the longitudinal motion controller. If the convergence process takes a relatively long time, the traffic mobility and roadway capacity are highly affected during this process. Specifically, if the consensus algorithm is applied to control the longitudinal motion of ramp merging vehicles, slow convergence rate also introduces safety issue since consensus must be reached before two vehicles merge with each other. Control gains should be set with the least time to consensus $\min t_{consensus}$ (they need to firstly satisfy constraint (24)). Consensus is reached when the following constraints are satisfied

$$\left| r_j\left(t_{consensus} - \tau_{ij}(t_{consensus})\right) - r_i(t_{consensus}) \right| \leq \eta_r \cdot$$
$$\left[ l_j + v_i(t_{consensus}) \cdot \left( t_{ij}^g(t_{consensus}) + \tau_{ij}(t_{consensus}) \right) \right] \quad (25)$$

$$\left| v_j\left(t_{consensus} - \tau_{ij}(t_{consensus})\right) - v_i(t_{consensus}) \right|$$
$$\leq \eta_v \cdot v_j\left(t_{consensus} - \tau_{ij}(t_{consensus})\right) \quad (26)$$

$$|a_i(t_{consensus})| \leq \delta_a \quad (27)$$

$$|jerk_i(t_{consensus})| \leq \delta_{jerk} \quad (28)$$

where $jerk_i$ is the derivative of vehicle $i$'s acceleration/deceleration, $\eta_r$ and $\eta_v$ are proportional thresholds of the headway consensus and speed consensus, respectively, $\delta_a$ and $\delta_{jerk}$ are differential thresholds of acceleration and jerk consensus, respectively.

*Constraint 3: Comfort Constraint.* The maximum changing rate of the consensus algorithm is defined as the maximum absolute value of acceleration/deceleration and jerk. This factor influences the ride comfort of the proposed longitudinal motion controller. A high maximum changing rate does not necessarily mean a high convergence rate, since algorithm (9) can either converge to consensus within a relatively short time but in a smooth manner, or converges to consensus within a relatively long time but change extremely fast at first. In this constraint, the maximum absolute value of acceleration/deceleration and jerk matter, since passengers on the vehicle would expect a comfort ride with acceleration/deceleration and jerk limited to certain intervals. The maximum changing rate of the consensus algorithm is evaluated by defining a parameter $\Omega$ as

$$\Omega_i = \omega_1 \cdot \max_{t \in [t_0, t_{consensus}]} (|a_i^{\max}(t)|, |d_i^{\max}(t)|)$$
$$+ \omega_2 \cdot \max_{t \in [t_0, t_{consensus}]} (|jerk_i^{\max}(t)|, |jerk_i^{\min}(t)|),$$
$$t \in [t_0, t_{consensus}] \quad (29)$$

where $a_i^{\max}$, $d_i^{\max}$, $jerk_i^{\max}$ and $jerk_i^{\min}$ denote the maximum acceleration, maximum deceleration, maximum jerk and minimum jerk of vehicle $i$, respectively, and $\omega_1$ and $\omega_2$ are weighting parameters. Control gains should be set with the minimum value of $\Omega$ in this constraint.

Upon proposing above three constraints, we propose *Algorithm 1* to build the 3-dimension lookup table for choosing the control gains. The set of $\Delta r_{ij}$ contains $\zeta 1$ elements, the set of $v_i$ contains $\zeta 2$ elements, and the set of $v_j$ contains $\zeta 3$ elements, therefore, the size of this lookup table is $\zeta 1 \times \zeta 2 \times \zeta 3$. These three sets $\Pi_{\Delta r_{ij}}, \Pi_{v_i}$ and $\Pi_{v_j}$ are sorted set with ascending order. Each combination of these three parameters maps to an ideal value of $k$ and an ideal value of $\gamma$, out of their sets $\Pi_\gamma$ and $\Pi_k$. Note that some specific initial condition cannot satisfy *Constraint 1*, as shown on line 04-05 of *Algorithm 1*. In that case, no value of control gains is generated, considering algorithm (9) being not functional under that particular condition.

```
Algorithm 1: Build lookup table offline
Input:  Π_Δr_ij = {Δr_ij_1, Δr_ij_2, …, Δr_ij_ζ1}, Π_v_i = {v_i_1, v_i_2, …, v_i_ζ2},
Π_v_j = {v_j_1, v_j_2, …, v_j_ζ3},  Π_γ = {v_γ_1, v_γ_2, …, v_γ_ζ4},  Π_k =
{v_k_1, v_k_2, …, v_k_ζ5}, ζ1 = |Π_Δr_ij|, ζ2 = |Π_v_i|, ζ3 = |Π_v_j|
Output: 3-dimension table with size ζ1 × ζ2 × ζ3
01: for ξ1 ∈ [1,ζ1], ξ2 ∈ [1,ζ2], ξ3 ∈ [1,ζ3], Δr_ij_ξ1 ∈ Π_Δr_ij,
        v_i_ξ2 ∈ Π_v_i, v_j_ξ3 ∈ Π_v_j do
02:     run algorithm (9) with Δr_ij_ξ1, v_i_ξ2 and v_j_ξ3
03:     find Λ_γ ⊆ Π_γ, Λ_k ⊆ Π_k satisfy Constraint 1
04:     if Λ_γ = ∅ || Λ_k = ∅ then
05:         γ_(ξ1,ξ2,ξ3) = NaN, k_(ξ1,ξ2,ξ3) = NaN
06:     else
07:         find Ψ_γ ⊆ Λ_γ, Ψ_k ⊆ Λ_k satisfy Constraint 2
08:         if |Ψ_γ| == |Ψ_k| == 1 then
09:             γ_(ξ1,ξ2,ξ3) ∈ Ψ_γ, k_(ξ1,ξ2,ξ3) ∈ Ψ_k
10:         else
11:             find Φ_γ ⊆ Ψ_γ, Φ_k ⊆ Φ_k satisfy Constraint 3
12:             if |Φ_γ| == |Φ_k| == 1 then
13:                 γ_(ξ1,ξ2,ξ3) ∈ Φ_γ, k_(ξ1,ξ2,ξ3) ∈ Φ_k
14:             else
15:                 γ_(ξ1,ξ2,ξ3) = min_{γ∈Φ_γ} Φ_γ, k_(ξ1,ξ2,ξ3) = min_{k∈Φ_k} Φ_k
16:             end if
17:         end if
18:     end if
19: end for
```

Once the lookup table has been generated, the initial condition of vehicles does not always match certain values while implementing algorithm (9)

$$\left(\Delta r_{ij}(t_0), v_i(t_0), v_j\left(t_0 - \tau_{ij}(t_0)\right)\right) \neq \left(\Delta r_{ij_{\xi_1}}, v_{i_{\xi_2}}, v_{j_{\xi_3}}\right) \quad (30)$$

Therefore, in order to find the ideal values of control gains given different initial conditions, the *Algorithm 2* is proposed. If the initial states of vehicles fall out of the ranges, invalid values are returned as shown on line 02-03, and algorithm (9) will not be applied to control the vehicle. In such cases, a backup safety-guarantee motion control algorithm will be applied to control the longitudinal motion of the vehicles.

```
Algorithm 2: Search lookup table in real time
Input:  (Δr_ij(t_0), v_i(t_0), v_j(t_0 − τ_ij(t_0))), (ζ1 × ζ2 × ζ3) size
of lookup table
Output: values of control gains k and γ
01: for Δr_ij(t_0), v_i(t_0), v_j(t_0 − τ_ij(t_0)) do
02:     if Δr_ij(t_0) < Δr_ij_1 || Δr_ij(t_0) > Δr_ij_ζ1 || v_i(t_0) < v_i_1 || v_i(t_0)
        > v_i_ζ2 || v_j(t_0 − τ_ij(t_0)) < v_j_1 || v_j(t_0 − τ_ij(t_0)) > v_j_ζ3 then
03:         return γ = NaN, k = NaN
04:     else
05:         find (Δr_ij_ξ1 | min |Δr_ij(t_0) − Δr_ij_ξ1|),
              (v_i_ξ2 | min |v_i(t_0) − v_i_ξ2|),
              (v_j_ξ3 | min |v_j(t_0 − τ_ij(t_0)) − v_j_ξ3|)
06:         return γ = γ(Δr_ij_ξ1, v_i_ξ2, v_j_ξ3), k = k(Δr_ij_ξ1, v_i_ξ2, v_j_ξ3)
07:     end if
08: end for
```

## V. SIMULATION RESULTS

In this section, we conduct numerical simulations to verify the effectiveness of the proposed longitudinal motion controller for CAV in car-following scenarios, which are based on the following four scenarios with different initial conditions of vehicles. These four scenarios stand for different typical initial conditions of vehicles, where $\Delta r_{ij}(t_0)$ can be either positive or negative, and $v_i(t_0)$ can be either larger or smaller than $v_j\left(t_0 - \tau_{ij}(t_0)\right)$. As shown in TABLE I, the initial headway between the leading vehicle $j$ and the following vehicle $i$ is a negative value in scenario 3 and 4. Since the proposed longitudinal motion controller is not only for car-following cases on the same lane (e.g., platooning), where the following vehicle is physically behind the leading vehicle, it can also be applied to ramp merging or intersection crossing cases, where the leading vehicle can be projected on the same lane as the following vehicle. Therefore, the initial distance headway between them can be negative.

TABLE I. Settings of Simulation Scenarios

|  | $\Delta r_{ij}(t_0)$ (m) | $v_i(t_0)$ (m/s) | $v_j\left(t_0 - \tau_{ij}(t_0)\right)$ (m/s) |
|---|---|---|---|
| Scenario 1 | 50 | 28 | 14 |
| Scenario 2 | 20 | 16 | 22 |
| Scenario 3 | -30 | 18 | 10 |
| Scenario 4 | -80 | 4 | 21 |

The parameters of the lookup table are set as: $\Pi_{\Delta r_{ij}} = \{-100, -90, \ldots, 100\}$ (m), $\Pi_{v_i} = \{2, 4, \ldots, 34\}$ (m/s), $\Pi_{v_j} = \{2, 4, \ldots, 34\}$ (m/s), $\Pi_\gamma = \{1, 2, \ldots, 10\}$, $k = 0.1$, $\zeta 1 = 21$, $\zeta 2 = 17, \zeta 3 = 17$. The threshold parameters in *Constraint 3* are set as $\eta_r = \eta_v = 0.05, \delta_a = 0.001, \delta_{jerk} = 0.005$. For the sake of simplicity while simulating algorithm (9), we assume the communication delay $\tau_{ij}(t)$ is a constant value of 60 ms [18]. The length of a vehicle is set as $l_j = 5$ m, and the desired time gap $t_{ij}^g(t)$ is set as a constant value of 0.7 s. We compare the proposed algorithm (9) with respect to the widely cited CACC algorithm proposed by van Arem *et al.* [19], and also with the authors' prior work Wang *et al.* [9].

The speed trajectories of the leading vehicle and the following vehicle under different scenarios are shown in Fig. 2, where each scenario has three different speed trajectories of the following vehicle's speed $v_i(t)$. The simulation results in terms of efficiency and comfort are shown in TABLE II. Since all three algorithms satisfy the safety constraint in these scenarios, the headway overshoot results are not shown here. Compared to the consensus algorithm the authors developed previously, the proposed algorithm in this study intensively decreases convergence time and maximum jerk. When compared to the consensus algorithm proposed by van Arem *et al.*, the convergence time is also reduced in all four scenarios. Although the proposed algorithm introduces a relatively higher maximum jerk in scenario 2 and 4, they are still in the comfort zone of a human passenger, which is [-10 m/s³, 10 m/s³] [20], [21].

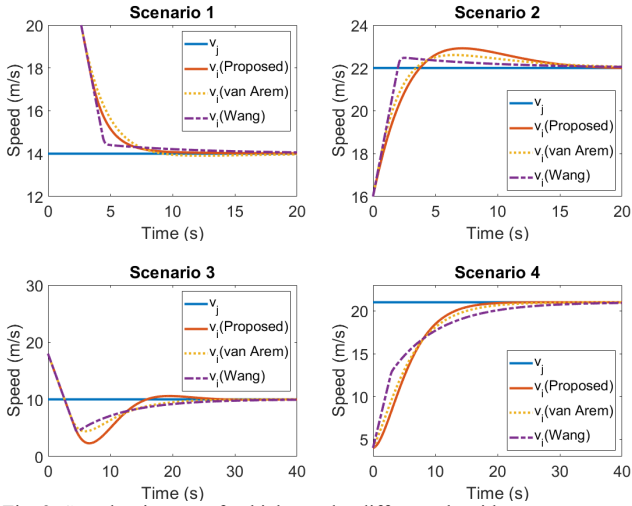

Fig. 2. Speed trajectory of vehicles under different algorithms

TABLE II. SIMULATION RESULTS

|  | Convergence time (s) | | | | Maximum jerk (m/s³) | | | |
|---|---|---|---|---|---|---|---|---|
| Scenario | 1 | 2 | 3 | 4 | 1 | 2 | 3 | 4 |
| Wang | 35.9 | 35.0 | 56.5 | 57.6 | 21.2 | 20.7 | 25.7 | 13.4 |
| van Arem | 29.3 | 32.1 | 41.8 | 40.1 | 1.5 | 1.6 | 2.3 | 0.7 |
| Proposed | 24.9 | 22.9 | 32.1 | 28.3 | 2.3 | 0.8 | 1.6 | 1.6 |

It should be noted that *Constraint 2* has a higher priority than *Constraint 3* in the proposed *Algorithm 2*, since the efficiency of the algorithm also relates to the ride safety to some extent. For example, in scenario 2 and 4, the initial longitudinal position of the following vehicle is in front of the leading vehicle. If these two vehicles cannot converge to position consensus (i.e., the following vehicle is behind the leading vehicle) before they enter the conflict zone of a ramp merging or intersection crossing scenario, potential crash might occur. Therefore, the proposed algorithm outperforms the other two algorithms due to its reduction of convergence time in all these four scenarios.

## VI. CONCLUSION

This paper proposed a consensus algorithm to control the real-time longitudinal motion of connected and automated vehicles. We developed an algorithm to build a lookup table for control gains in the proposed longitudinal controller offline, and another algorithm to search the ideal values of control gains based on different initial conditions online. Numerical simulation showed the proposed algorithm outperforms the authors' prior work, as well as van Arem's linear feedback-based CACC algorithm in all four different scenarios in terms of convergence time and maximum jerk of the simulation run.

One future direction of this work is to improve the algorithms to build a more comprehensive lookup table, especially when leading vehicle's speed is dynamically changing. Also, higher-order vehicle dynamics can be considered to extend the current double-integrator consensus algorithm, where the acceleration and jerk of the leading vehicle can be included to generate a more accurate reference control signal for the following vehicle.